\documentstyle[12pt]{article}
\def\strut{\rule[-.5cm]{0cm}{1cm}}
\def\sspace{\baselineskip = .16in}

\begin{document}
\title{From Fermion Mass Matrices to Neutrino
Oscillations\thanks{Supported in part by Department of Energy Grant
\#DE-FG02-91ER406267}}

\author{{\bf K.S. Babu}\\ and\\ {\bf Q. Shafi}\\ Bartol Research Institute\\
University of Delaware\\ Newark, DE 19716}

\date {}
\maketitle
\begin{abstract}

We present an ansatz for the quark and lepton mass matrices,
derivable from SO(10) type GUTs, which
accommodates a heavy $(> 92 GeV)$ top quark and permits large mixings
in the $\nu_\mu \leftrightarrow \nu_\tau$ sector (as suggested by
the recent Kamiokande and IMB data on the atmospheric neutrinos). The
well known asymptotic relations $m_b = m_\tau$, $m_s = \frac{1}{3}
m_\mu$ and $m_dm_s = m_e m_\mu$ all hold to a good approximation.
Depending on $\nu_\mu \leftrightarrow \nu_\tau$ mixing which can
even be maximal, the mixing angle relevant for solar neutrino oscillation
lies in the range $7.8 \times 10^{-3} \stackrel{_<}{_\sim} \sin^2
2\theta_{e\mu} \stackrel{_<}{_\sim} 2.1 \times 10^{-2}$. For the
$^{71}$Ga experiment the event rate, normalized against the standard
solar model prediction of 132 SNU, is estimated to be between 80 and
20 SNU.

\end{abstract}
\newpage
\sspace

There is currently a surge of experimental activity around the
exciting possibility that one or more species of the neutrinos may
possess a tiny mass. The most recent analysis$^{(1)}$ of the
Kamiokande$^{(2)}$ and the IMB data$^{(3)}$  on atmospheric
neutrino interactions suggest a
discrepancy with theoretical expectations$^{(4)}$ which can be nicely
explained in terms of neutrino oscillations$^{(1,5)}$. Assuming a
two flavor
$(\nu_\mu \leftrightarrow \nu_\tau)$ oscillation, the relevant mixing
angle satisfies the constraint $\sin^22\theta_{\mu\tau}
\stackrel{_>}{_\sim} 0.42$. The mass difference squared lies
in the range $(10^{-1}-10^{-3}) eV^2$.

Perhaps the most surprising aspect of this observation, if true, is
the relatively large value suggested for the $\nu_\mu-\nu_\tau$
mixing angle $\theta_{\mu\tau} (\stackrel{_>}{_\sim} 20^\circ)$.
Within the framework of grand unified theories with the minimal higgs
structure, for instance, one typically finds that$^{(6,7)}$
$\theta_{\mu\tau}
\simeq \mid V_{cb}\mid \simeq 0.05$, where $V_{cb}$ is the $(cb)$
element of the well known KM matrix of the quark sector. By making
some minor modifications in the higgs structure, one could perhaps
make $\theta_{\mu\tau}$ three times larger than $\mid V_{cb}\mid$,
but this still is much too small compared to what the atmospheric
neutrino data suggests.

The purpose of this paper is to address this and some important
related issues within a more general framework, which is inspired by GUTs as
well as some earlier work of Fritzsch$^{(8)}$ and others. The idea is to
write down an ansatz for the fermion mass matrices which i) is
predictive in the quark sector and admits a heavy $(> 92 GeV)$ top quark,
ii) preserves to a good approximation some well known asymptotic relations
including $m_b \simeq m_\tau, m_s \simeq \frac{1}{3} m_\mu, m_dm_s
\simeq m_em_\mu$ and $\theta_c \simeq (m_d/m_s)^\frac{1}{2}$, and
iii) allows for large mixings in the $\nu_\mu \leftrightarrow \nu_\tau$
sector. It turns out that in the quark sector the new ansatz we propose
has essentially the same predictive capacity as the original one of ref. (8).
A particularly important feature of the new scenario is the restriction on
the mixing angle relevant for the       solar neutrino problem. One finds a
significant deficit relative to the standard solar model prediction,
which will soon be tested in the ongoing SAGE$^{(9)}$ and GALLEX experiments.

The ansatz that we will consider can be motivated within a grand unified
framework (e.g., SO(10))$^{(10)}$ with a non-minimal higgs structure. It works
both with supersymmetric (SUSY) as well as with non-SUSY GUTs. In the
latter case, in order for the proton lifetime to be compatible with
the experimental lower bound, an intermediate step (e.g., $SU(4)_c
\times SU(2)_L \times SU(2)_R$)$^{(11)}$ would be needed. Consider then the
following ansatz for the quark and lepton mass matrices:

$$
M_d = \left( \begin{array}{ccc}
0 & A & 0\\
A & De^{i\alpha} & B\\
0 & B & C\end{array}
\right)\hspace{.5in}
M_u = \left( \begin{array}{ccc}
0 & A^\prime & 0\\
A^\prime & 0 & B^\prime\\
0 & B^\prime & C^\prime\end{array}
\right)$$

$$
M_l = \left( \begin{array}{ccc}
0 & A & 0\\
A & -3De^{i\alpha} & -3B\\
0 & -3B & C\end{array}
\right)\hspace{.3in}
M_\nu^{Dirac} = \left( \begin{array}{ccc}
0 & A^\prime & 0\\
A^\prime & 0 & -3B^\prime\\
0 & -3B^\prime & C^\prime\end{array}
\right)$$

\begin{equation}
M_\nu^{Majorana} = \left( \begin{array}{ccc}
M_1e^{i\gamma_{1}} & 0 & 0\\
0 & M_2e^{i\gamma_{2}} & M_3e^{i\gamma_{3}}\\
0 & M_3e^{i\gamma_{3}} & 0\end{array}
\right).
\end{equation}

Several comments are in order:

\begin{enumerate}
\item We have written down the matrices of Eq. (1) after a
suitable redefinition of the phases of the fermion fields. The
non-zero elements of the matrices, before this redefinition, are
allowed to have arbitrary phases. All but one phase $\alpha$ can be
removed from the Dirac mass matrices. The charged current interaction
in this basis is not proportional to the identity matrix, but has the
generation structure given in terms of two phase parameters $\sigma$
and $\tau$ by

\begin{equation}
\left( \begin{array}{ccc}
1 & &\\
& e^{i\sigma} &\\
& & e^{i\tau}\end{array}
\right).
\end{equation}\nonumber

\noindent
The entries $A, A^{\prime}, B, B^{\prime}$,... can be chosen to be real and
positive without loss of generality.
\item Within an SO(10) type framework, the entries $A, A^{\prime},
C, C^{\prime}$ arise from the higgs {\bf 10} plets, while $B,
B^\prime$ and
$D$ arise from the {\bf 126} plets. The reason for the form of the
$D$ terms is to retain, to a good approximation, the well-known
asymptotic relation $m_s \simeq \frac{1}{3} m_\mu$ $^{(12)}$.
The $B,B^\prime$
terms are motivated by the desire to accommodate both large mixings in the
2-3 lepton sector and a moderately heavy top quark $(m_t
\stackrel{_<}{_\sim} 160 GeV)$.  When the parameter $B$ is set to zero,
our ansatz reduces to that of Ref. 12, a detailed analysis of which
is presented in Ref. 13;
\item With $C \gg A,B,D$, one would recover $m_b = m_\tau$ $^{(14)}$ to
a good
approximation. However, a small violation induced through mixing of this
asymptotic equality plays an important role in our analysis;
\item The form of the 2-3 sector of the heavy Majorana mass matrix
is directly related to the corresponding sectors of the quark and the
charged lepton matrices. In other words, the same {\bf 126} plet of
higgs contributes to these three sectors.  The (22) elements $D$ and $M_2$
of Eq. (1) arise from the same Yukawa coupling to a {\bf 126} of Higgs.
Similarly, $B, B^{\prime}$ and $M_3$ result from a common Yukawa coupling
to another Higgs {\bf 126} plet.  (Note that the {\bf 126} generating
(22) elements $D$ and $M_2$ should be distinct from the {\bf 126} that
generates
(23) elemnt $B, B^{\prime}$ and $M_3$, if it were the same, a non-zero (33)
element will also be induced.  Consequently, there is no relation of the
type $M_2/M_3 = D/B$.)
We need an additional
(independent) contribution in the Majorana sector (the 1-1 entry) to
give a large mass to the heavy $\nu_{eR}$;
\item Since the charged fermion sector is described by 11 fundamental
parameters including tan$\beta$, the ratio of the
two Higgs vacuum expectation values, we will have 3 predictions.
Additionally, since some of these parameters are phase angles
whose moduli are constrained to be less than unity, we also have approximate
relations among the physical observables (see below).
\end{enumerate}

\noindent
It should be emphasized that the above form of the mass matrices is prescribed
at the
GUT scale $M_U$.

The real symmetric matrix $M_u$ is diagonalized by an orthogonal
transformation, $\tilde{O}_u M_u O_u = M_u^{diag}$ (tilde denotes
transpose), where

\begin{equation}
O_u \simeq \left( \begin{array}{ccc}
1 & - \sqrt{\frac{m_u}{m_c}} & \sqrt{\frac{m_u}{m_t}}
\frac{m_c}{m_t}\strut\\
\sqrt{\frac{m_u}{m_c}} & 1 & \sqrt{\frac{m_c}{m_t}}\strut\\
- \sqrt{\frac{m_u}{m_t}} & - \sqrt{\frac{m_c}{m_t}} & 1\end{array}
\right).
\end{equation}\nonumber

\noindent
Here $m_u, -m_c$ and $m_t$ denote the eigenvalues of $M_u$ and we
have used the relations $A^\prime \simeq \sqrt{m_um_c}, B^\prime
\simeq \sqrt{m_cm_t}$ and $C^\prime \simeq m_t$.

Since $M_d$ and $M_\ell$ are complex symmetric matrices, their
diagonalization is achieved via bi-unitary
transformations of the type $\tilde{U}_d M_d U_d = M_d^{diag}$, and
$\tilde{U}_\ell M_\ell U_\ell = M_\ell^{diag}$. Denoting the eigenvalues
as $(m_d, -m_s, m_b)$ and
$(m_e, - m_\mu, m_\tau)$ respectively, the appropriate unitary
matrices are:

\begin{equation}
U_d \simeq \left( \begin{array}{ccc}
e^{-i\phi/2} & - \sqrt{\frac{m_d}{m_s}} e^{-i\phi/2} &
\epsilon \sqrt{\frac{m_d}{m_b}} \sqrt{\frac{m_s}{m_b}}\strut\\
\sqrt{\frac{m_d}{m_s}} e^{i\phi/2} & e^{i\phi/2} &
\epsilon\strut\\
-\epsilon\sqrt{\frac{m_d}{m_s}} &       -\epsilon & 1\end{array}
\right),
\end{equation}\nonumber
\vspace{.2in}

\begin{equation}
U_l \simeq \left( \begin{array}{ccc}
e^{-i\tilde{\phi}/2} & - \sqrt{\frac{m_e}{m_\mu}}
e^{-i\tilde{\phi}/2} & -3\epsilon \sqrt{\frac{m_e}{m_\tau}}
\sqrt{\frac{m_\mu}{m_\tau}}\strut\\
\sqrt{\frac{m_e}{m_\mu}} e^{i\tilde{\phi}/2} & e^{i\tilde{\phi}/2} &
-3\epsilon\strut\\
3\epsilon \sqrt{\frac{m_e}{m_\mu}} &
3\epsilon e^{i\tilde{\phi}/2} & 1\end{array}
\right).
\end{equation}\nonumber

\noindent
Here we have defined

\begin{equation}
\epsilon \equiv \frac{1}{4} \sqrt{1-\frac{m_b^2}{m_\tau^2}}~,
\end{equation}

\noindent
assumed $C \gg B,D \gg A$, and used the relations

\begin{equation}
\begin{array}{ccl}
A & \simeq & \sqrt{m_dm_s} \simeq \sqrt{m_em_\mu}\nonumber\strut\\
B & \simeq & \epsilon m_b\nonumber\strut\\
C & \simeq & m_b \simeq m_\tau\nonumber\strut\\
D & = & \frac{m_b}{2} \left[ \left(3\frac{m_s^2}{m_b^2} + \frac{1}{9}
\frac{m_\mu^2}{m_b^2}\right)
-3\epsilon^4\right]^{\frac{1}{2}}\nonumber\strut\\
D\cos\alpha & = & \pm m_b \left[ \epsilon^2 +
\frac{\epsilon^{-2}}{8} \left(\frac{m_s^2}{m_b^2} -
\frac{1}{9}\frac{m_\mu^2}{m_b^2}\right)\right].
\end{array}
\end{equation}

\noindent
The phases $\phi, \tilde{\phi}$ are determined in terms of B, C, D
and $\alpha$ through the relations

\begin{equation}
\begin{array}{lcl}
(B^2-CDe^{i\alpha}) & = & \mid B^2-CDe^{i\alpha}\mid
e^{-i\phi},\nonumber\strut\\
(B^2+\frac{1}{3}CDe^{i\alpha}) & = & \mid
B^2+\frac{1}{3}CDe^{i\alpha}\mid e^{-i\tilde{\phi}}.\end{array}
\end{equation}

\noindent
{}From the relation $D^2 \geq 0$, one finds that

\begin{equation}
\frac{3m_s^2}{m_b^2} + \frac{1}{9} \frac{m_\mu^2}{m_b^2} \geq
3\epsilon^4.
\end{equation}

\noindent
Employing $\mid\cos\alpha\mid \leq 1$, one can show that

\begin{equation}
\frac{1}{2}\sqrt{\left| \mid\frac{m_s}{m_b}\mid -
\frac{1}{3}\mid \frac{m_\mu}{m_b}\mid \right|} \leq \epsilon
\leq \frac{1}{2}\sqrt{\mid\frac{m_s}{m_b}\mid +
\frac{1}{3}\mid\frac{m_\mu}{m_b}\mid}.
\end{equation}

\noindent
We therefore see that in the limit of a strict equality $m_b =
m_\tau$, we have $\mid m_\mu\mid = 3\mid m_s\mid$. In practice,
$\epsilon$ will be a small positive quantity which would lead to
small violations of these relations.

In the quark sector the Kobayashi-Maskawa matrix is given by

\begin{equation}
V_{KM} = \tilde{O}_u \left( \begin{array}{ccc}
1 & &\nonumber\\
& e^{i\sigma} &\nonumber\\
& & e^{i\tau}\end{array} \right) U^*_d
\end{equation}

\noindent
which leads to the following (asymptotic!) matrix elements:

\begin{equation}
\begin{array}{ccl}
\mid V_{us}\mid & = & \mid V_{cd}\mid\; =\; \mid \sqrt{\frac{m_d}{m_s}} -
\sqrt{\frac{m_u}{m_c}} e^{i(\sigma - \phi)} \mid \nonumber\strut\\
\mid V_{ub}\mid & = & \mid \epsilon\sqrt{\frac{m_d}{m_b}}
\sqrt{\frac{m_s}{m_b}} + \sqrt{\frac{m_u}{m_c}} e^{i\sigma} \left(
\epsilon - \sqrt{\frac{m_c}{m_t}} e^{i(\tau -
\sigma)}\right) \mid\nonumber\strut\\
\mid V_{cb}\mid & = & \mid V_{ts}\mid\; =\; \mid \epsilon -
\sqrt{\frac{m_c}{m_t}} e^{i(\tau - \sigma)}\mid \nonumber\strut\\
\mid V_{td}\mid & = & \mid \sqrt{\frac{m_u}{m_t}} \frac{m_c}{m_t} +
\sqrt{\frac{m_d}{m_s}} e^{i(\sigma - \phi)} \left(
\sqrt{\frac{m_c}{m_t}} - \epsilon e^{i(\tau -
\sigma)}\right) \mid.\end{array}
\end{equation}

\noindent
In addition, the reparameterization invariant CP violating quantity J
is given by$^{(15)}$

\begin{equation}
\begin{array}{ccl}
J & = & Im \left( V_{us} V_{cb} V^*_{ub} V^*_{cs} \right)\strut\\
& \simeq & [ \epsilon \sqrt{\frac{m_u}{m_t}}
\sqrt{\frac{m_d}{m_s}} [ \sin (\phi - \tau) + \sin \left( \phi + \tau
- 2 \sigma \right) ]\strut\\
& - & \sqrt{\frac{m_u}{m_t}} \sqrt{\frac{m_c}{m_t}}
\sqrt{\frac{m_d}{m_s}} \sin \left( \phi - \sigma \right) -
\epsilon^2 \sqrt{\frac{m_u}{m_c}} \sqrt{\frac{m_d}{m_s}}
 \sin \left( \phi - \sigma \right) ]. \end{array}
\end{equation}

It is clear from Eqs. (12-13) that once the charged fermion masses are
specified, all the mixing angles as well as $J$ are determined in
terms of the phases $\sigma$ and $\tau$. This situation is analogous
to the Fritzsch ansatz, except that our modified version can
accommodate a heavy top quark (due to the difference in the expression
for $|V_{cb}|$).

As was pointed out earlier, the above relations hold at the GUT scale
and so comparison with data requires that we consider their evolution
with momenta. It was noted quite some time ago$^{(16)}$ that as far as the
mixing angles are concerned, significant evolution will occur for the
entries involving the third family provided that the top quark is
sufficiently heavy $(\geq 100 GeV)$.

For definiteness, we focus on the supersymmetric case. We further
simplify our analysis by assuming that the sparticles and the second higgs
doublet
are degenerate at 300 GeV. Employing the
one loop renormalization group equations, and fixing the top
quark mass at $130 GeV$, we first determine the unification
scale $M_U$. As low energy inputs we use the precisely known values
of $\alpha_1$ and $\alpha_2$ at $M_Z^{(17)}$: $\alpha_1(M_Z) = 0.01013,
\alpha_2(M_Z) = 0.03322$. We find the
unification scale to be $M_U = 9.8 \times 10^{15} GeV$ and the common
gauge coupling at $M_U$ to be $\alpha_U(M_U) = \frac{1}{25.8}$.
Running backwards, we determine $\alpha_3 (M_Z)$ to be 0.105,
which is in agreement with measurements.

The evolution equations for the elements of the KM matrix can be
found in ref. (16). It turns out that for $\tan \beta
\stackrel{_<}{_\sim} 7 (\tan \beta \equiv v_2/v_1$, the well-known
ratio of the two vevs present in the minimal SUSY extension of the
standard model), the coupled equations can be solved semi-analytically.
Here we will content ourselves by presenting results (Table 1)
showing the variation of the mixing angles with momentum, for varying
values of $m_t$ and $\tan\beta$. For $m_t$ not too large
$(\stackrel{_<}{_\sim} 130 GeV)$, the variation in $\mid V_{cb}\mid,
\mid V_{ub}\mid, \mid V_{td}\mid, \mid V_{ts}\mid$ is
$\stackrel{_<}{_\sim} 5\%$.
However, if the top is close to its maximal allowed value of about
$190 GeV$, the variations in some of the quantities can exceed 20\%.
In Table 2 we display the dependence of the quark and charged
lepton mass ratios $m(M_U)/m(\mu)$, on $m_t$ and $\tan \beta$.
For the $d$ and $s$ quarks, as well as
for the $(e, \mu, \tau)$ leptons these ratios are
(essentially) independent of $m_t$ and $\tan\beta$ and are in the
range (0.177-0.179) and (0.656-0.689) respectively.

{}From Tables 1-2, we can infer the predictions of our ansatz for the
quark masses and mixing angles at the weak scale. From the inequality
of Eq. (10) we see that the parameter $\epsilon$ is bounded by
$\epsilon \leq 0.10$. There is also a lower bound on $\epsilon,
\epsilon \geq 0.035$, if we assume that the top mass is around
$130~ GeV$ (see Eq. (12) for $|V_{cb}|$).  This lower bound, however,
goes away if the top quark is much heavier, near its fixed
point value.$^{13,19}$
If $\epsilon$ is set
to zero, one recovers the asymptotic
relation$^{(12,13,18,19)}$

\begin{equation}
\mid V_{cb}\mid \simeq (m_c/m_t)^{\frac{1}{2}}.
\end{equation}

\noindent
This relation (Eq. (14)) is disfavored phenomenologically
(although not excluded), both in the non-supersymmetric as well as in the
supersymmetric case. It
requires a top quark in the mass range $180-220 GeV (175-190 GeV)$ in
the non-SUSY (SUSY) case. Moreover $\mid V_{cb}\mid$ must be
$\geq 0.052^{13}$ to be compared with the experimentally favored range
$0.043 \pm 0.009$.

{}From the constraint $0.035 \leq \epsilon \leq 0.10$ one can deduce
the allowed range for the $b$-quark mass.  It follows from Table 2
that $m_b(m_b)$ lies in the range $(4.3 - 5.3) GeV$ (for $\alpha_s(M_Z)
=0.105$), the larger value
corresponding to smaller $\epsilon$.  This is in accord with the value
quoted in Ref. 20, $m_b(m_b) = 4.25 \pm 0.1 ~GeV$.
(We should point out that $m_b$ can be brought further down, as in Ref. 13,
even if $\epsilon=0$, provided the top quark mass is around 180 GeV.  Here,
however, we are interested in a moderately heavy top ($m_t \sim 130~GeV$).)
Since the asymptotic relation
$m_b = m_\tau$ holds to within 10\%, it follows from Eq. (10) that
$\mid m_s\mid = \frac{1}{3} \mid m_\mu\mid$ should hold to within 5\%.
This implies
$m_s (1 GeV) = 130-140 MeV$, in relatively good agreement$^{(20)}$
with observations. From the asymptotic relation $m_dm_s = m_em_\mu^{(12)}$,
we also have a prediction for the $d$ quark mass: $m_d = (5.7-6) MeV$.

Since the variation of the Cabibbo angle with momentum is negligible,
we essentially have the Fritzsch prediction for $\mid V_{us}\mid$. On
the other hand, $\mid V_{ub}\mid$ can increase by as much as 14\%
(even for $m_t < 160 GeV)$ so that at the weak scale, $\mid
V_{ub}\mid = 0.002-0.0037$. The CP parameter $J$ naturally comes out
in the range $10^{-4}-10^{-5}$ consistent with observations in the
$K$ meson system.

To summarize, in the charged fermion sector, our ansatz is quite
predictive. It reproduces the well-known asymptotic relations $m_b =
m_\tau, m_s = \frac{1}{3} m_\mu$ and $m_dm_s = m_em_\mu$ to a good
approximation. The KM angles in the quark sector are determined in
terms of the fermion masses and two arbitrary phase angles.
Specifically, there are three predictions in the charged fermion sector.
One of them is the mass relation $m_dm_s=m_em_{\mu}$, the other two are
mixing angle relations of Eq. (12) obtained by eliminating the phases
$\sigma$ and $\tau$.  The
predictive capacity of our ansatz is as good as Fritzsch's for
quark mixing angles, but the top quark can be heavy in our case.

Turning now to the neutral lepton sector, the $3 \times 3$ light
neutrino mass matrix is obtained from the see-saw formula$^{(21)}$

\begin{equation}
M_\nu = M_\nu^{Dirac} (M_\nu^{Majorana})^{-1} \tilde{M}_\nu^{Dirac}
\end{equation}

\noindent
where

\begin{equation}
(M_\nu^{Majorana})^{-1} \equiv M^{-1} \left( \begin{array}{ccc}
r_1e^{i\phi_{1}} & 0 & 0\strut\\
0 & 0 & r_2e^{i\phi_{2}}\strut\\
0 & r_2e^{i\phi_{2}} & 1\end{array} \right).
\end{equation}

\noindent
[Note that we
are making
the assumption that the direct contributions to the left handed
Majorana neutrino masses are negligible. This is justified in a SUSY SO(10)
framework, since the tree-level vacuum expectation value of the field
which supplies such Majorana masses can consistently be set to
zero.]  Here $M$ is an overall superheavy scale, and $r_1, r_2$ denote ratios
of superheavy masses.  The parameter $r_2$ is given by $r_2=M_3/M_2$.
It is reasonable to assume $r_2$ to be of order 1, since $M_3/M_2 \sim
B/D$ which is of order 1, (see eq. (7)) times ratio of vev's.  We will
also discuss the case when $r_2$ is not of order 1.
 The matrix $M_\nu$ takes the form

\begin{equation}
M_\nu = \frac{m_t^2}{M} P \left[ \begin{array}{ccc}
0 & -3 \sqrt{\frac{m_u}{m_t}} \frac{m_c}{m_t}r_{2} &
\sqrt{\frac{m_u}{m_t}} \sqrt{\frac{m_c}{m_t}}r_2\strut\\
-3 \sqrt{\frac{m_u}{m_t}} \frac{m_c}{m_t} r_{2} & 9
\frac{m_c}{m_t} + \frac{m_u}{m_t} \frac{m_c}{m_t} r_1e^{i\phi_{1}} &
-3\sqrt{\frac{m_c}{m_t}} + 9 \frac{m_c}{m_t} r_2 e^{i\phi_{2}}\strut\\
\sqrt{\frac{m_u}{m_t}} \sqrt{\frac{m_c}{m_t}} r_2 &
-3\sqrt{\frac{m_c}{m_t}} + 9 \frac{m_c}{m_t} r_2 e^{i\phi_{2}} &
1-6 \sqrt{\frac{m_c}{m_t}} r_2 e^{i\phi_{2}}\end{array} \right] P
\end{equation}

\noindent
where $P$ is the diagonal phase matrix $P = diag [e^{i\phi_{2}}, 1,
1]$.  Observe that $M_{\nu}$ has five parameters,
viz., $r_1, r_2, \phi_1, \phi_2$ and
$M$.  Note however, that the phase $\phi_1$ appears with a very small
coefficient making it an irrelevant parameter.  The four effective
parameters describe six observables: three neutrino masses and the
three neutrino mixing angles.  This
leads to two predictions.  We shall see below that
these two are predictions for mixing angles in terms of
the neutrino mass ratios.

It is convenient to diagonalize $M_\nu$ in two stages. Let
${M}_\nu^{\prime}$ denote the part of $M_\nu$ which excludes the two
diagonal phase matrices. We can reduce ${M}_\nu^{\prime}$ to an
effective $2 \times 2$ matrix (with zeroes along the first row and
column) through the transformation $\tilde{U}_1 {M}_\nu^{\prime} U_1$,
where $U_1$ is a unitary matrix given by

\begin{equation}
U_1 = \left( \begin{array}{ccc}
e^{i\phi_{2}} & -\frac{1}{9} \sqrt{\frac{m_u}{m_c}} &
-\frac{1}{3}\sqrt{\frac{m_u}{m_t}}\strut\\
\frac{1}{9}\sqrt{\frac{m_u}{m_c}} & e^{-i\phi_{2}} & 0\strut\\
\frac{1}{3}\sqrt{\frac{m_u}{m_t}} & 0 & e^{-i\phi_{2}}\end{array}
\right).
\end{equation}

\noindent
Note that we are justified in neglecting the term $\frac{m_u}{m_t}
\frac{m_c}{m_t} r_1 e^{i\phi_{1}}$ as long as $r_1$ is of order unity.

The 2-3 sector of the transformed matrix coincides with that of
${M}_\nu^{\prime}$. It can be diagonalized by a unitary matrix $U_2$ of
the type

\begin{equation}
U_2 = e^{i\gamma} \left( \begin{array}{ccc}
1 & 0 & 0\strut\\
0 & ce^{i(\rho + \delta )} & se^{i(\rho - \delta )}\strut\\
0 & -se^{-i(\rho - \delta )} & ce^{-i (\rho + \delta )}\end{array}
\right)
\end{equation}

\noindent
where $c \equiv \cos \theta$ and $s \equiv \sin \theta$. In
particular, the expressions for $\theta$ and $\rho$ are as follows:

\begin{equation}
\tan 2 \theta = \frac{6 \sqrt{\frac{m_c}{m_t}} \mid 3
\sqrt{\frac{m_c}{m_t}} (1-3 \sqrt{\frac{m_c}{m_t}} r_2
e^{-i\phi_{2}}) + (1-3 \sqrt{\frac{m_c}{m_t}} r_2
e^{+i\phi_{2}}) (1-6 \sqrt{\frac{m_c}{m_t}} r_2
e^{-i\phi_{2}}) \mid}
{\left[\left| 1-6 \sqrt{\frac{m_c}{m_t}} r_2 e^{i\phi_{2}}\right|^2 -
(9 \frac{m_c}{m_t})^2\right]}
\end{equation}

\begin{equation}
\tan 2 \rho = \frac{-3 \sqrt{\frac{m_c}{m_t}} r_2 \sin \phi_2 \left(
1 + 9 \sqrt{\frac{m_c}{m_t}} \frac{m_c}{m_t} \right)}{\left[ 1 - 9
\sqrt{\frac{m_c}{m_t}} \cos \phi_2 + 9 \frac{m_c}{m_t} + 18
\frac{m_c}{m_t} r_2^2 - 27 \frac{m_c^2}{m_t^2} \cos \phi_2 \right]}
\end{equation}

The light neutrino mass eigenvalues (mass eigenstates are denoted
$\nu_{1,2,3}$) turn out to be:

\begin{equation}
\begin{array}{lcl}
m_{\nu_{1}} & = & \frac{m_u^2}{81 M} r_1\strut\\
m_{\nu_{2}} & = & 81 \frac{m_c^2}{M} r_2^2 \left[ \mid 1 - 6
\sqrt{\frac{m_c}{m_t}} r_2 e^{i\phi_{2}} \mid^2 + 2 \mid -3
\sqrt{\frac{m_c}{m_t}} + 9 \frac{m_c}{m_t} r_2 e^{i\phi_{2}}\mid^2 +
\left( 9 \frac{m_c}{m_t} \right)^2 \right] ^{-\frac{1}{2}}\strut\\
m_{\nu_{3}} & = & \frac{m_t^2}{M} \left[ \mid 1 - 6
\sqrt{\frac{m_c}{m_t}} r_2 e^{i\phi_{2}} \mid^2 + 2 \mid -3
\sqrt{\frac{m_c}{m_t}} + 9 \frac{m_c}{m_t} r_2 e^{i\phi_{2}}\mid^2 +
\left( 9 \frac{m_c}{m_t} \right)^2 \right]^{\frac{1}{2}}.\end{array}
\end{equation}

We are now in a position to write down the elements of the lepton KM
matrix $V_{KM}^{lepton}$. Let $U_\nu \equiv U_1 U_2$. Then

\begin{equation}
V_{KM}^{lepton} = U_\nu^\dagger \left( \begin{array}{ccc}
e^{i\phi_{2}} & &\\
& e^{i\sigma} &\\
& & e^{i\tau}\end{array} \right) U_l
\end{equation}

\noindent
and one obtains for the off-diagonal elements the expressions:

\begin{equation}
\begin{array}{lcl}
\left| V_{\nu_{1}\mu}\right| & = & \left| \sqrt{\frac{m_e}{m_\mu}} -
\frac{1}{9} \sqrt{\frac{m_u}{m_c}} e ^{i(\sigma - \tilde{\phi})}
\right|\nonumber\strut\\
\left| V_{\nu_{1}\tau} \right| & = & \left| \frac{1}{3}
\sqrt{\frac{m_u}{m_c}} e^{i\sigma} \left( \epsilon -
e^{i(\tau - \sigma)} \sqrt{\frac{m_c}{m_t}} \right) +
3\epsilon \sqrt{\frac{m_e}{m_\tau}}
\sqrt{\frac{m_\mu}{m_\tau}} \right|\nonumber\strut\\
\left| V_{\nu_{2}e} \right| & = & \cos \theta \left|
\sqrt{\frac{m_e}{m_\mu}} - \frac{1}{9} \sqrt{\frac{m_u}{m_c}}
e^{-i(\sigma - \tilde{\phi})} \right| \nonumber\strut\\
\left| V_{\nu_{2}\tau} \right| & = & \left| V_{\nu_{3}\mu} \right| =\left|
3\epsilon \cos \theta +
\sin \theta e^{i(\tau - \sigma + 2\rho)} \right| \nonumber\strut\\
\left| V_{\nu_{3}e} \right| & = & \left| \sin \theta \left(
\sqrt{\frac{m_e}{m_{\mu}}} - \frac{1}{9} \sqrt{\frac{m_u}{m_c}}
e^{i(\tilde{\phi}-\sigma)} \right) + 3\epsilon \cos \theta
\sqrt{\frac{m_e}{m_\mu}}
e^{i(\frac{\tilde{\phi}}{2}-\sigma-2\phi_{2})} \right. \nonumber\strut\\
& & \left. -\frac{1}{3} \cos \theta \sqrt{\frac{m_u}{m_t}} e^{i(2\rho
+\tilde{\phi})} \right|. \end{array}
\end{equation}

Several comments are in order:

\begin{enumerate}
\item Unlike the quark sector, the mixing angles in the lepton sector
do not run between $M_U$ and the weak scale
since the right-handed neutrinos are superheavy.
\item The mass ratio $m_{\nu_{2}}/m_{\nu_{3}}$ is enhanced by
almost two orders of magnitude relative to what one naively expects
from the see-saw mechanism. This is again related to the factor 3 in
the ansatz for the fermion mass matrices which, in turn, was
motivated by the desire to include large mixings. The enhancement
will be important when we discuss atmospheric and solar neutrino
oscillations.
\item The mixing angle relevant for the MSW explanation of the
solar neutrino puzzle is given by

\begin{equation}
\mid V_{\nu_{2}e}\mid = \cos \theta \left| \sqrt{\frac{m_e}{m_\mu}} -
\frac{1}{9}\sqrt{\frac{m_u}{m_c}} e^{i(\sigma-\tilde{\phi})} \right|.
\end{equation}

\noindent
Note that as $\theta \rightarrow 0$ the expression for $\mid
V_{\nu_{2}e}\mid$ essentially
coincides with the one derived earlier in ref. (6). If, the
recent Kamiokande results on the atmospheric neutrino survive the
test of time, then $\theta$ is expected to lie in the range $18^\circ
\stackrel{_<}{_\sim} \theta \stackrel{_<}{_\sim} 45^\circ$ (see
below). Taking all this into account, we estimate that $7.8 \times
10^{-3} \stackrel{_<}{_\sim}
\sin^2 2\theta_{e \mu} \stackrel{_<}{_\sim} 2.1 \times 10^{-2}$. The
corresponding $\mid \Delta m^2\mid$ is in the range $(5 \times
10^{-6} - 2 \times 10^{-6}) eV^2$. For
the Gallium experiments currently under way the estimated event rate$^{(22)}$,
normalized relative to the standard solar model value of 132 SNU,
lies between 80 and 20 SNU.

\item A combined fit to the solar neutrino and the atmospheric
neutrino data requires $m_{\nu_{2}} = (1.4-2.2) \times 10^{-3} eV,
m_{\nu_{3}} = (0.03-0.6) eV$. The ratio $m_{\nu_{2}}/m_{\nu_{3}}$ should
then lie in the range $(2.3 \times 10^{-3} - 7.3 \times 10^{-2})$.
{}From Eq. (22), this constraint implies a lower limit on
the parameter $r_2$, $r_2 > 0.8$.  If we restrict to $r_2 = (0.8-3)$ so
that there is no hierarchy in the superheavy masses, the 2-3
mixing angle $\theta$ in the neutrino sector
is in the range $\theta = (18^\circ-45^\circ)$.
This certainly is in
the right parameter range for atmospheric neutrino oscillations.
For large values of $r_2$,
the angle $\theta$ approaches $3/2\sqrt{m_c/m_t} \sim 7^\circ$
(see eq. (20)).  Even
in this case, the atmospheric neutrino deficit can be accommodated, if
$\epsilon$ is near its upper limit of 0.1 (see $V_{\nu_2 \tau}$ of eq. (24)).
Note that $\theta$ close to 45$^0$ gives the largest count ($\sim$ 80 SNU)
for the SAGE/GALLEX experiment.  The effective
$\nu_{\mu} \leftrightarrow \nu_{\tau}$ mixing angle for this case
lies in the range
$0.69 \stackrel{_<}{_\sim} {\rm sin}^22\theta_{\mu \tau}
\stackrel{_<}{_\sim} 1$.
\end{enumerate}

In conclusion, the scheme presented here was guided by the desire to
retain to the extent possible the simplicity of the original Fritzsch
ansatz and come up with some predictions in the neutrino sector. The
SAGE/GALLEX experiment should soon be able to tell if we are on the
right track!

\section*{Note added:}  After the submission of this paper for publication,
GALLEX collaboration has announced their result: $83 \pm 19 \pm 8$ SNU.
This is in remarkable agreement with our prediction (see eq. (25) and
the subsequent discussions).
\newpage

\section*{References}

\begin{enumerate}
\item E.W. Beier et al., University of Pennsylvania preprint (1992).
\item K.S. Hirata et al., Kamiokande-II Collaboration, submitted to
Phys. Lett. B (1992).
\item R. Becker-Szendy et al., Boston University preprint BUHEP-91-24
(1992).
\item T.K. Gaisser, Talk given at the Fermilab Workshop on the
Many Aspects of Neutrino Physics, November 1991, Bartol Preprint
BA-92-03 (1992) and references therein.
\item For earlier phenomenological analysis see:\\
J. Learned, S. Pakvasa and T. Weiler, Phys. Lett. \underline{B207},
79 (1988);\\
V. Barger and K. Whisnant, Phys. Lett. \underline{B209}, 365 (1988);\\
K. Hidaka, M. Honda, and S. Midorikawa, Phys. Rev. Lett.
\underline{61}, 1537 (1988);\\
C. Albright, Phys. Rev. \underline{D45}, 725 (1992).
\item G. Lazarides and Q. Shafi, Nucl. Phys. \underline{B350}, 179 (1991);\\
ibid. \underline{B364}, 3 (1991).
\item S. Bludman, D. Kennedy and P. Langacker, Phys. Rev. D
\underline{45}, 1810 (1992).
\item H. Fritzsch, Phys. Lett. \underline{B70}, 436 (1977);\\
ibid. \underline{73B}, 317 (1978).
\item A.I. Abazov et al. Phys. Rev. Lett. \underline{67} 3332 (1991).
\item H. Georgi, in ``Particles and Fields,'' C.E. Carlson, ed., AIP,
New York (1974)\\
H. Fritzsch and P. Minkowski, Ann. Phys. \underline{93}, 193 (1975).
\item J.C. Pati and A. Salam, Phys. Rev. \underline{D10}, 275 (1974).
\item H. Georgi and C. Jarlskog, Phys. Lett., \underline{B89}, 297
(1979).
\item S. Dimopoulos, L. Hall and S. Raby, Phys. Rev. Lett. \underline{68},
1984 (1992);\\
V. Barger, M.S. Berger, T. Han and M. Zralek, University of Wisconsin
preprint MAD/PH/693 (1992).
\item M. Chanowitz, J. Ellis and M.K. Gaillard, Nucl. Phys.
\underline{B128}, 506 (1977);\\
A. Buras, J. Ellis, M.K. Gaillard and D. Nanopoulos, Nucl. Phys.
\underline{B135}, 66 (1978).
\item C. Jarlskog, Phys. Rev. Lett., \underline{55}, 1039 (1985).
\item K.S. Babu, Z. Phys. C \underline{35}, 69 (1987);\\
K. Sasaki, Z. Phys. C \underline{32}, 149 (1986);\\
B. Grzadkowski, M. Lindner and S. Theisen, Phys. Lett. \underline{B198},
64 (1987).
\item U. Amaldi, Wim de Boer, and H. F\"{u}rstenau, Phys. Lett.
\underline{260B}, 447 (1991).
\item M. Freire, G. Lazarides and Q. Shafi, Mod. Phys. Lett.
\underline{A5}, 2453 (1991).
\item J. Harvey, P. Ramond and D. Reiss, Nucl. Phys.
\underline{B199}, 223 (1982);\\
P. Ramond, University of Florida preprint 92-4 (1992);\\
\item J. Gasser and H. Leutwyler, Phys. Rep. \underline{87}, 77
(1982).
\item M. Gell-Mann, P. Ramond and R. Slansky, in Supergravity, ed.,
F. van Nieuwenhuizen and D. Freedman (North Holland, 1979), p. 315;\\
T. Yanagida, in "Proceedings of the Workshop on Unified Theory and
Baryon Number in the Universe", ed. A. Sawada and H. Sugawara,
(KEK, Tsukuba, Japan, 1979).
\item See for example:\\
V. Barger, R.J.N. Phillips and K. Whisnant, Phys. Rev. D
\underline{43}, 1110 (1991);\\
A.J. Baltz and J. Weneeser, Phys. Rev. Lett. \underline{66}, 520
(1991).
\end{enumerate}
\newpage

\noindent
{\large\bf Table 1:} The evolution factor $\left|
\frac{V_{ij}(m_t)}{V_{ij}(m_U)} \right|$ for the quark mixing angles
$\mid V_{ub}\mid$, $\mid V_{cb}\mid$, $\mid V_{td}\mid$ and
$\mid V_{ts}\mid$ (as functions of $m_t$ (in GeV) and
$\tan\beta$).  The corresponding factor for the CP paramaeter
$J$ is given by the square of these numbers.
The running of all the other
elements of $V$ are negligible. The dashes indicate that the top
quark Yukawa coupling has become non-perturbative before the GUT
scale.
\vspace{.2in}

$$\begin{array}{|c|c|c|c|c|}\hline
& & & &\\
m_t\rightarrow & 100 & 130 & 160 & 190\\
\hrulefill & & & &\\
\downarrow\!\tan\!\beta & & & &\\\hline
1 & 1.051 & 1.139 & -- & --\\\hline
2 & 1.027 & 1.055 & 1.122 & --\\\hline
3 & 1.023 & 1.046 & 1.093 & 1.359\\\hline
4 & 1.022 & 1.043 & 1.085 & 1.244\\\hline
5 & 1.021 & 1.041 & 1.081 & 1.217\\\hline
6 & 1.021 & 1.041 & 1.080 & 1.205\\\hline
7 & 1.021 & 1.040 & 1.079 & 1.198\\\hline
\end{array}
$$
\newpage
\noindent
{\large\bf Table 2:} The evolution factors $\left[
\frac{m_i(M_U)}{m_i(\mu)} \right]$ for the quark masses as function
of $m_t$ and $\tan\beta$. For $u$ quark, $\mu = 1 GeV$ is chosen
and for $c$ and $b$ quarks, $\mu = 1.27 GeV$ and $4.25 GeV$
respectively. The corresponding factors for $(d,s)$ and $(e, \mu,
\tau)$ are independent of $m_t$ and $\tan\beta$ and are equal to
0.178 and 0.673 respectively.
\vspace{.2in}

$$\begin{array}{|c|c|c|c|c|}\hline
& & & &\\
m_t\rightarrow & 100 & 130 & 160 & 190\\
\hrulefill & & & &\\
\downarrow\!\tan\!\beta\hfill & & & &\\\hline
1\hspace{.2in} \begin{array}{cc}u\\c\\b\end{array} &
\begin{array}{cc}0.201\\0.212\\0.242\end{array} & \begin{array}{cc}
0.257\\0.272\\0.263\end{array} & \begin{array}{cc} --\\--\\--\end{array} &
\begin{array}{cc} --\\--\\--\end{array}\\\hline
2\hspace{.2in} \begin{array}{cc}u\\ c\\b\end{array} &
\begin{array}{cc}0.187\\ 0.198\\0.236\end{array} &
\begin{array}{cc}0.205\\
0.216\\0.244\end{array} & \begin{array}{cc}0.249\\ 0.263\\0.260\end{array} &
\begin{array}{cc}--\\ --\\--\end{array}\\\hline
3\hspace{.2in} \begin{array}{cc}u\\c\\b\end{array} &
\begin{array}{cc}0.185\\ 0.196\\0.236\end{array} &
\begin{array}{cc}0.199\\
0.211\\0.242\end{array} & \begin{array}{cc}0.230\\ 0.243\\0.254\end{array} &
\begin{array}{cc}0.448\\ 0.474\\0.317\end{array}\\\hline
4\hspace{.2in} \begin{array}{cc}u\\c\\b\end{array} &
\begin{array}{cc}0.184\\ 0.195\\0.235\end{array} &
\begin{array}{cc}0.198\\
0.209\\0.241\end{array} & \begin{array}{cc}0.225\\ 0.238\\0.252\end{array} &
\begin{array}{cc} 0.343\\ 0.363\\0.290\end{array}\\\hline
5\hspace{.2in} \begin{array}{cc}u\\ c\\b\end{array} &
\begin{array}{cc}0.184\\ 0.194\\0.235\end{array} & \begin{array}{cc}
0.197\\ 0.208\\0.241\end{array} & \begin{array}{cc} 0.223\\
0.236\\0.251\end{array} & \begin{array}{cc}0.322\\
0.340\\0.284\end{array}\\\hline
6\hspace{.2in} \begin{array}{cc} u\\ c\\b \end{array} &
\begin{array}{cc}0.184\\ 0.194\\0.235\end{array} & \begin{array}{cc}
0.196\\
0.208\\0.241\end{array} & \begin{array}{cc} 0.222\\ 0.234\\0.251\end{array} &
\begin{array}{cc} 0.312\\ 0.330\\0.281\end{array}\\\hline
7\hspace{.2in} \begin{array}{cc} u\\ c\\b\end{array} &
\begin{array}{cc} 0.184\\ 0.194\\0.235\end{array} &
\begin{array}{cc}0.196\\ 0.207\\0.240\end{array} & \begin{array}{cc}
0.221\\ 0.234\\0.250\end{array} & \begin{array}{cc} 0.307\\
0.325\\0.280\end{array}\\\hline
\end{array}
$$

\end{document}